\documentclass[longauth]{aa}  
\usepackage{graphicx}
\usepackage{caption}
\usepackage{subcaption}
\usepackage{xcolor}
\usepackage{soul}
\usepackage{dblfloatfix}
\usepackage{txfonts}
\usepackage{mathabx}
\usepackage[hidelinks]{hyperref}
\usepackage[normalem]{ulem}
\usepackage{orcidlink}
\usepackage{booktabs}
\usepackage{verbatim}
\usepackage{siunitx}
\usepackage{textcomp}
\usepackage{amsmath}
\usepackage{amsfonts}
\usepackage{amssymb}
\usepackage{float}

\newcounter{pta}

\DeclareMathSizes{12}{30}{16}{12}
\usepackage{amsmath,scalerel}

\makeatletter
\renewcommand*\aa@pageof{, page \thepage{} of \pageref*{LastPage}}
\makeatother

\usepackage{natbib}
\setcitestyle{authoryear,open={(},close={)}} 

\usepackage{titlesec}

\titlespacing\section{0pt}{14pt plus 2pt minus 2pt}{5pt plus 1pt minus 1pt}
\titlespacing\subsection{0pt}{12pt plus 4pt minus 2pt}{6pt plus 2pt minus 2pt}
\titlespacing\subsubsection{0pt}{12pt plus 4pt minus 2pt}{6pt plus 2pt minus 2pt}

\usepackage[switch]{lineno} 

\usepackage{upgreek}

\begin{document}

\title{Long-term Timing Results of Ecliptic Pulsars Observed with I-LOFAR}

\author{
S.~C.~Susarla\orcidlink{0000-0003-4332-8201}\inst{\ref{uog}},
O.~A.~Johnson\orcidlink{0000-0002-5927-0481}\inst{\ref{tcd},\ref{ucb}}, D.~J.~McKenna\orcidlink{0000-0001-7185-1310}\inst{\ref{astron}}, E.~F.~Keane\orcidlink{0000-0002-4553-655X}\inst{\ref{tcd}}, P.~J.~McCauley\orcidlink{0000-0003-4399-2233}\inst{\ref{tcd}},
J.~P.~W.~Verbiest\orcidlink{0000-0002-4088-896X}\inst{\ref{FSI}},
C.~Tiburzi\orcidlink{0000-0001-6651-4811}\inst{\ref{inaf-oac}},
A.~Golden\orcidlink{0000-0001-8208-4292}\inst{\ref{uog}}}
\institute{
Physics, School of Natural Sciences \& Center for Astronomy, College of Science and Engineering, University of Galway, University Road, Galway, H91TK33, Ireland.\label{uog}\and
School of Physics, Trinity College Dublin, College Green, Dublin 2, D02 PN40, Ireland\label{tcd}\and
Radio Astronomy Laboratory, University of California, Berkeley, CA, USA\label{ucb}\and
ASTRON, The Netherlands Institute for Radio Astronomy, Oude Hoogeveensedijk 4, 7991 PD Dwingeloo, The Netherlands\label{astron}\and
Florida Space Institute, University of Central Florida, 12354 Research Parkway, Partnership 1 Building, Suite 214, Orlando, 32826-0650, FL, USA\label{FSI}\and
INAF - Osservatorio Astronomico di Cagliari, via della Scienza 5, 09047 Selargius (CA), Italy\label{inaf-oac}
}

   \date{Received XXX; accepted YYY}
\authorrunning{LOFAR}

 
  \abstract
{Pulsar timing at low frequencies offers a powerful tool for studying the interstellar medium. Additionally, pulsar observations in the ecliptic enables us to study the effects of the solar wind which becomes much more prominent at low radio frequencies. The Irish station of the LOw Frequency ARray~(I-LOFAR) is a sensitive low-frequency radio telescope, capable of delivering high-precision data for pulsar studies.}
{We present a comprehensive dataset of times-of-arrival, timing solutions and dispersion measure (DM) time series for seven ecliptic pulsars observed over two-to-three years with I-LOFAR. The primary objectives are to investigate time-dependent dispersion effects and provide high-precision timing data for pulsar timing experiments.}
{We measure DM variations through pulsar timing and analysed these across different ecliptic latitudes to assess the impact of the solar wind on each pulsar. We model the intrinsic pulse-profile variability as a function of frequency.}
{The high-precision DM time series for all seven pulsars exhibit clear variations dependent on their ecliptic latitudes, revealing the impact of the solar wind. Some pulsars show significant changes in their pulse widths across the frequency band, while others remain stable. We examine and quantify the pulse-nulling present in PSR~J0826$+$2637, we report evidence for DM chromaticity in PSR~J1645$-$0317, and we describe how PSR~J2145$-$0750's DM precision is such that it could resolve the ionospheric DM contribution. This makes it a target of interest for telescopes in areas of the globe where the ionospheric electron density is higher, e.g. the Murchison Radio Observatory in Australia.}
{This data release underscores the potential of I-LOFAR, or any standalone international LOFAR station, for advancing low-frequency pulsar studies, particularly in analyses of dispersion in the interstellar medium, the solar wind and the ionosphere.}


   \maketitle



\section{Introduction}
Pulsars are highly magnetized, fast-spinning neutron stars, remnants of massive stars following a supernova~(\citealt{gold1968,LorimerandKramer2005}). Pulsars emit beams of electromagnetic radiation from their polar regions~\citep{ckf99,ml99,psc15}; and as their magnetic and rotational axes are in general misaligned, these beams sweep across space while the star rotates. When one of the beams cuts through the line-of-sight (LoS) of an observer from Earth, regular pulses of radiation are observed, which are particularly prominent at radio frequencies. These pulses, when folded over the rotational period, form a characteristic pulse profile that is stable over time~(\citealt{helfand1975,Kuo2012}). This makes pulsars ideal for a number of experiments, based on the technique of pulsar \textit{timing}~(see e.g.~\citealt{stairs2003} and references therein). Pulsar timing is a process that entails the recording of the times-of-arrival (ToAs) of pulses emitted from a given pulsar~\citep{edwards2006}. ToAs are derived by cross-correlating a pulsar signal integrated over many rotation periods, typically $10^4$ to $10^5$, with a high signal-to-noise (S/N) template~\citep{taylor1992}. 
Given the stability requirement, ``high-precision pulsar timing'' (with a residual root-mean-square, rms, lower than 1$\mu$s) is achieved by long-term monitoring of millisecond pulsars (MSPs;~see~\citealt{mspcitation}), rather than the majority of pulsars which rotate with periods of order of half a second or more. 
Several phenomena introduce noise into the timing 
model (see~\citet{verbiest2018} for a review). Among these noise sources are gravitational waves (GWs), which can be systematically identified by examining correlated patterns in the residuals across an array of pulsars~\citep{hd1983}; 
this approach underlies the concept of pulsar timing array (PTA) experiments.
The current state-of-the-art~\citep{verbiest2024} has seen hints of such a correlated signature in data from various PTA consortia at a significance level from $3$ to $4.2\sigma$~\citep{eptagwb,nanogravgwb,pptagwb,CPTA2024,agazie2024}.
In identifying any GW signal in PTA datasets, one of the most challenging signatures to disentangle is that of the ionised interstellar medium (IISM). The IISM noise process can dominate the GW background if not modelled properly, but its inverse dependency on the squared of the observing frequency offers a leverage to implement effective mitigation techniques, especially thanks to the usage of low radio-frequency pulsar data. Indeed, one of the best ways to characterise the IISM noise process is to observe pulsars with facilities that reach radio-frequencies $<1$~GHz, and large fractional bandwidths~\citep{verbiest2018}. 

This paper presents a study of seven pulsars observed between $100$ and $200$~MHz, with high fractional bandwidth over the last two-to-three years, using the Irish station of the Low Frequency Array (LOFAR; see \citealt{vanhaarlem2013}), hereafter referred to as I-LOFAR (also known as IE613). These seven pulsars are located near the ecliptic plane with various ecliptic latitudes (ELAT), so that their signal traverses the solar wind once per year during their Solar conjunctions, offering an excellent means of assessing the solar contribution to the IISM signature. The data products developed as part of this work are provided open source~\citep{susarla_2025_dr}\footnote{\texttt{https://doi.org/10.5281/zenodo.14865192}}. This paper describes these data and our pipelines; and illustrates some of the enabled scientific investigations. In \S~\ref{dacol} we outline the dataset and our data collection methods, including choices made for the observational settings and the initial processing steps. In \S~\ref{proarch} we describe the post-processing techniques applied to the data. 
\S~\ref{results} presents the results and discusses their implications; finally, \S~\ref{conclusions} concludes with a summary of the findings and future research directions.

\begin{figure*}[h]
    \centering
    \includegraphics[width=0.8\linewidth]{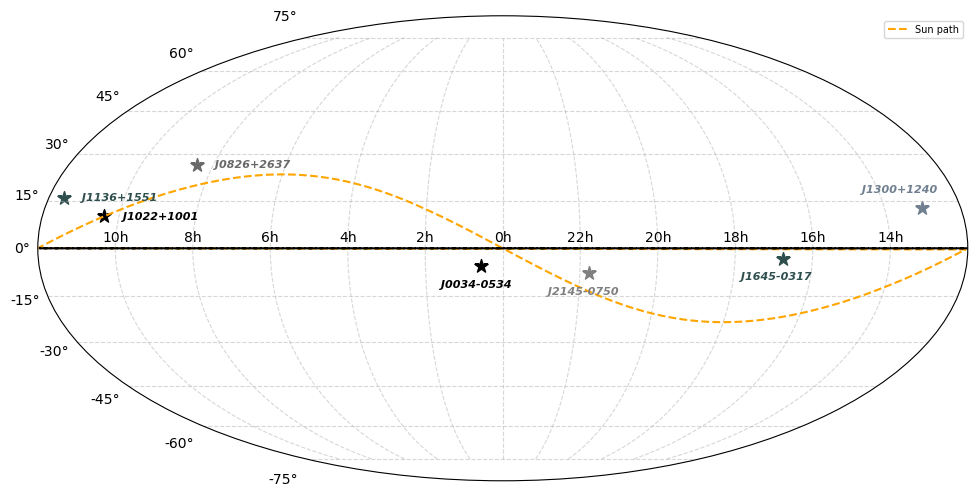}
    \caption{An Aitoff projection of the locations of the pulsars listed in Tab.~\ref{tab:pulsar_table}. The pulsars are shown in star symbols as well as the ecliptic plane, which is indicated by a yellow dashed line.}
    \label{fig:skymap}
\end{figure*}

\section{I-LOFAR: Telescope and Observing campaign}
\label{dacol}

In this section, we describe the Irish international LOFAR (I-LOFAR) telescope used for this study and its configurations for pulsar data collection. We also describe the selection criteria that led to the choice of the observed pulsars.
\subsection{I-LOFAR Telescope}
LOFAR is a pan-European radio telescope array consisting of $52$ \textit{stations} from as far West as Ireland to as far East as Latvia. The \textit{core} of the array is in the Netherlands where there are 24 core stations centred on a ``superterp'' near Exloo, the Netherlands. There are a further 14 \textit{remote} stations within the Netherlands and 14 \textit{international} stations. At present the whole array is being upgraded~\citep{lofar2023whitepaper} to, amongst other improvements, increase the instantaneous bandwidth and the clock distribution. Furthermore two new international stations, in Italy and in Bulgaria, will soon join the array, significantly improving its North-South extent. The dataset collected with the Irish LOFAR station, is the subject of this paper. I-LOFAR is located at the Rosse Observatory\footnote{\url{https://www.rosseobservatory.ie/}} in Birr, Co. Offaly, in the centre of the island of Ireland. During the observing campaign, I-LOFAR operated in standalone mode for typically $31$ hours per week and it was within this allocation, over the past three years that data were collected for the presented project. The I-LOFAR high-band antennas (HBA) were used, observing between $100$ and $200$~MHz. For each observation the HBA were analogue-beamformed in the direction of the target with the resultant data sampled using a $200$-MHz clock, resulting in one real data sample per antenna polarisation every $5$~ns. A coarse channelisation is performed using a polyphase filterbank which creates $512$ channels of complex voltage data per polarisation every $5.12\;\upmu$s. The hardware limits the accessible data to $488$ of these $512$ sub-bands for 8-bit recording. The data in the vicinity of the FM band contain no useful information due to a hardware filter, thus we choose to include in the final observation sub-bands 12--499, that correspond to a lowest and highest topocentric frequencies of about 102.2461 and 197.5586 MHz.

\subsection{Pulsar selection}
\label{pulsel}
Since its inception, I-LOFAR has been conducting pulsar observations. For this study, we focused on a carefully selected sample of seven pulsars observed consistently over a two-to-three year period, each satisfying the condition $|\textrm{ELAT}| < 20^{\circ}$. These pulsars were chosen based on the list published in \citet{Tiburzi2021}, ensuring a wide temporal coverage of solar conjunctions throughout the year. Our aim was twofold: to enhance the cadence of the low-frequency dataset contributed to the International PTA (IPTA) database \citep{Perera2019} for the PTA pulsars and to ensure a robust data coverage for studying solar wind (SW) effects. To finalize the sample, we observed all pulsars listed in \citet{Tiburzi2021} at least once to determine optimal integration times while adhering to the telescope's operational limit of 60 minutes per observation. Given the weekly constraint of a maximum observing time of 31 hours, we selected the seven best sources, optimizing the sample based on available time and ensuring a spread in Right Ascensions. Out of the seven selected sources, two are part of PTA samples.

Fig.~\ref{fig:skymap} displays the location of these pulsars relative to the Sun's path over a year. Notably, each pulsar's LoS aligns closely with the Sun at different times of the year, depending on its Right Ascension. Tab.~\ref{tab:pulsar_table} provides a summary of key properties for these pulsars, including the median signal-to-noise ratio (S/N), which indicates their relative brightness in the LOFAR (HBA) frequency range.


\begin{table*}[ht] 
    \centering 
    \begin{tabular}{lccrrrrl} 
        \toprule
        Pulsar Name 
        & Time span, $t_{\rm span}$ 
        & \multicolumn{1}{c}{Average Integration} 
        & \multicolumn{1}{c}{Period} 
        & \multicolumn{1}{c}{Ecl.~Lat.} 
        & \multicolumn{1}{c}{DM}
        & \multicolumn{1}{c}{Median}\\
        (J2000) 
        & (MJD) (yr)
        & \multicolumn{1}{c}{time (min)} 
        & \multicolumn{1}{c}{(ms)} 
        & \multicolumn{1}{c}{(deg)} 
        & \multicolumn{1}{c}{(pc/cm$^3$)}
        & \multicolumn{1}{c}{S/N}\\
        \midrule 
        J0034$-$0534 & 59892$-$60431 (2.0 yr)& 50 & 1.8 & $-8.53$ & 13.7650  & 72.2\\
        J0826+2637 & 59695$-$60411 (2.5 yr)& 15 & 530.6 & $7.24$ & 19.5041  & 522.5\\
        J1022+1001 & 59803$-$60542 (2.0 yr)& 50 & 16.5 & $-0.06$ & 10.2532  & 40.6\\
        J1136+1551 & 59695$-$60430 (2.5 yr)& 15 & 1187.9 & $12.17$ & 4.8482  & 917.7\\
        J1300+1240 & 59695$-$60431 (2.5 yr)& 60 & 6.2 & $17.63$ & 10.1588  & 39.5\\
        J1645$-$0317 & 59696$-$60431 (2.5 yr)& 20 & 387.6 & $18.89$ & 35.7416  & 612.0\\
        J2145$-$0750 & 59696$-$60530 (3.0 yr)& 45 & 16.1 & $5.31$ & 9.0027  & 119.8\\ 
        \bottomrule
    \end{tabular}
    \centering
    \caption{Key properties of the observed ecliptic pulsars presented in this study. The table summarizes the pulsars' J2000 names, observation timespans in MJD and years, average integration times, rotational periods, ecliptic latitudes, fiducial DMs, and the median S/Ns over the entire observing span. }
    \label{tab:pulsar_table}
\end{table*}

\subsection{Dispersion Measure precision}\label{dm}
As an electromagnetic wave traverses the IISM, it undergoes \textit{dispersion} due to the free electrons present in it. This effect is characterised by the \textit{dispersion measure} (DM) parameter, defined as the total electron density along the line of sight between Earth and a pulsar, affecting the arrival times of radio pulses by causing frequency-dependent delays. The DM of a pulsar is given by the integral of the column density $n^{LoS}_e$ along the line of sight (LoS):
\begin{equation}
   \mathrm{DM} = \int_{\rm LoS} n^{LoS}_e {\rm d}l.
   \label{dmeq}
\end{equation}
The dispersion causes a delay in the arrival of pulsar emission that depends on both the frequency of the radiation and the DM parameter, expressed in units of $\textrm{pc cm}^{-3}$. This delay can be calculated as:
\begin{equation}
   \Delta t = \frac{\mathrm{DM}}{K_D \nu^2}\;,
   \label{timedel}
\end{equation}
where $K_{\rm D}$ is the dispersion constant, and $\nu$ is the observing frequency in MHz. The constant $K_{\textrm{D}}$ itself is a combination of physical constants, each measured to some finite level, and SI conventions. We note that in our work here we adopt a standard pulsar timing convention to fix this constant at the exact value of $K_{\rm D} \coloneqq 2.41 \times 10^{-4}$~$\textrm{MHz}^{-2} \textrm{pc cm}^{-3}\textrm{s}^{-1}$~\citep{LorimerandKramer2005}. This is consistent with the relevant pulsar software packages, namely, \textsc{DSPSR}~\citep{vanstraaten2011} and \textsc{tempo2}~\citep{Hobbsetal2006a}. As the dispersion delay is inversely proportional to the square of the frequency, it becomes significantly more pronounced at low frequencies. 
In this study, the average DM precision obtained on an individual observation across all pulsars is $1.2 \times 10^{-4}$~$\textrm{pc cm}^{-3}$.

\subsection{Observations and data analysis} \label{obs}
During the standalone time, regular observations of the seven chosen pulsars were conducted over the course of two-to-three years, accumulating a total of $\sim$240 hours of combined observations. Science-ready data products were produced using the REAL-time Transient Acquisition backend~\citep[REALTA;][]{REALTA_2021}. Baseband voltages were recorded to disk, and \textsc{udpPacketManager}~\citep{UDPDAVIDJOSS2023} was used to create intermediate data products in the \textsc{PSRDADA}\footnote{\texttt{https://psrdada.sourceforge.net/index.shtml}} format from these voltages. Furthermore, coherent dedispersion is applied to the baseband voltages using the \texttt{digifil} tool of \textsc{DSPSR}. Coherent dedispersion is a technique wherein an input signal is convolved with the inverse transfer function of the IISM, effectively removing the dispersion delay signature caused by free electrons along the line of sight~\citep{hankins1975}. Dual-polarisation complex products are produced for each observation, which were then converted into \textsc{SigProc}~\citep{sigproc} filterbank files using the \texttt{digifil} tool with a frequency resolution of $24.41~\textrm{kHz}$ and source-dependent temporal resolution of at least $81.92~\upmu\textrm{s}$. The filterbanks are then converted into TIMER \textit{archive} \citep{hotanpsrchive} files, a standard pulsar data format, by folding them using \texttt{dspsr}.


\begin{figure*}[!h]
    \centering
    \includegraphics[width=0.9\linewidth,height=1.02\linewidth]{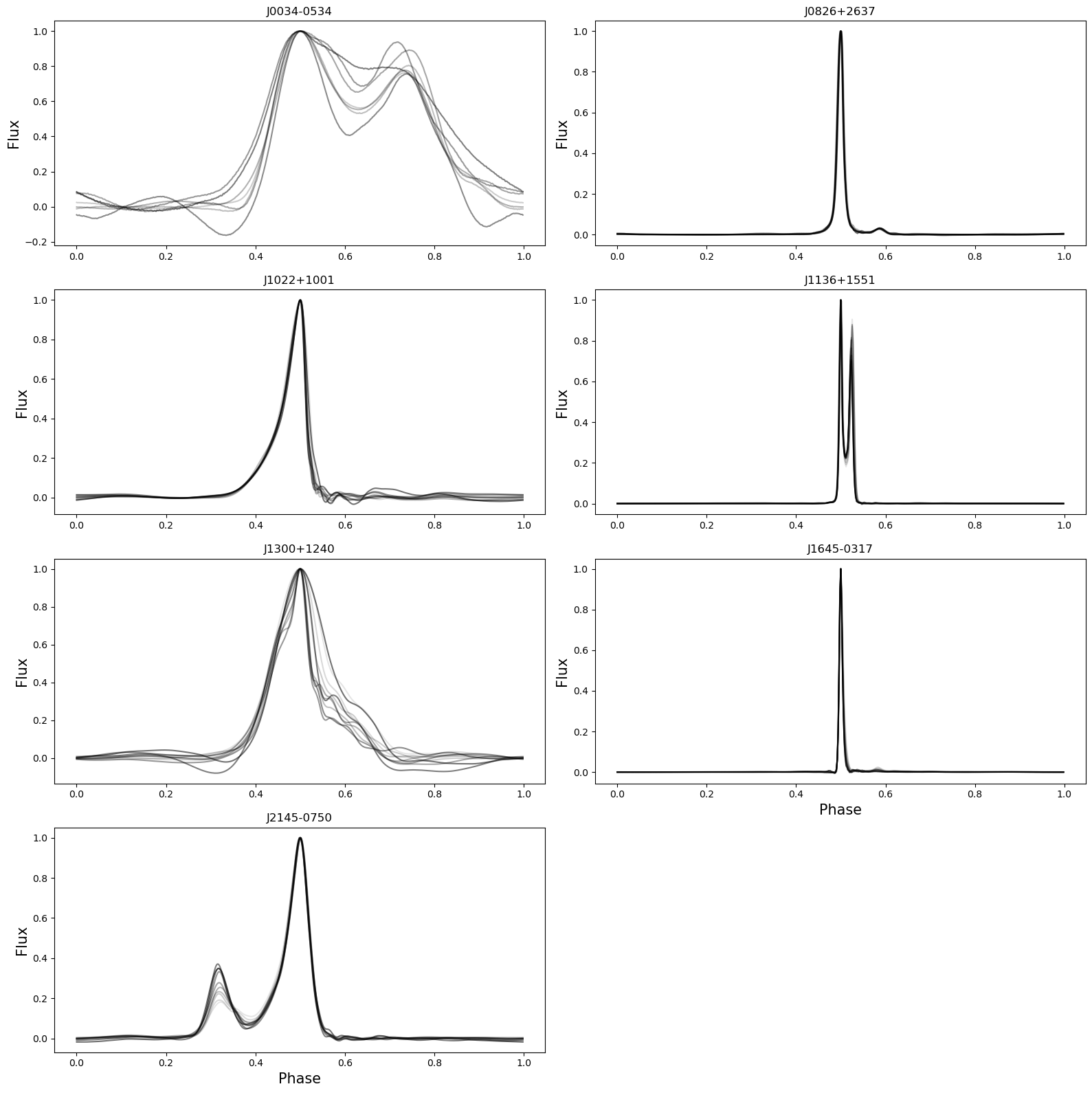}
    \caption{Frequency evolution of normalized pulse profiles. The colour intensity is scaled according to the observing frequency. Darker shades correspond to a high frequency and lighter shades correspond to a lower frequency.}
    \label{fig:freq_evol}
\end{figure*}


\section{Data processing}
\label{proarch}
The final data have a central frequency of 149.902~MHz and were coherently dedispersed to this frequency. Then, they were folded into 10~s sub-integrations, using the \textsc{DSPSR} software suite. Once the archives were created according to these specifications, several post-processing steps were applied. In the post-processing phase, each observation was cleaned of radio frequency interference (RFI) using a modified version of the \textsc{CoastGuard} software package (see \citet{lazaruscoast}, \href{https://www2.physik.uni-bielefeld.de/fileadmin/user_upload/radio_astronomy/Publications/Masterarbeit_LarsKuenkel.pdf}{Kuenkel, in prep.}\footnote{\url{https://github.com/larskuenkel/iterative_cleaner}}) and corrected for parallactic angle rotation and projection effects~\citep{jones1941,hamaker1996} with the \textsc{dreamBeam} software package\footnote{\url{https://github.com/2baOrNot2ba/dreamBeam}}. Each observation was then time-averaged and partially frequency-averaged to ten frequency sub-bands using the \textsc{PSRCHIVE} software suite \citep{hotanpsrchive,straatenpsrchive}. The final bandwidth was cut to range from 112 to 190 MHz, as a small portion of the wider band was affected by consistent RFI and the edges of the high/low pass filters present in the system.

\subsection{Creation of templates}
\label{cretempl}
All time-averaged archives, except those within 45 degrees of the Sun, are combined using the \texttt{psradd} tool from \textsc{PSRCHIVE} to make one individual archive file. For PSR~J0034$-$0534, given its complexity, we selected the highest S/N observation instead of adding all the observations. We then frequency-average the channels down to 10, and smoothened such data-derived \textit{templates} with the wavelet-smoothing technique offered by the \texttt{psrsmooth} tool. This process is performed iteratively after optimizing the pulsar model (or \textit{ephemeris}) via the timing technique (see \S~\ref{timing}), updating the archives' ephemerides to ensure accurate DM values in the profiles. The presence of any residual dispersion in the final template is checked using the \texttt{pazi} module, with corrections applied to the DM as needed. The result is a high-S/N, noiseless template~(see Fig.~\ref{fig:freq_evol}).

\subsection{Pulsar timing}
\label{timing}
The procedure described in the previous section is used to create a preliminary template from all of the cleaned archives. ToAs were then created for each observation (that was time-averaged and frequency-averaged down to 10 channels) and channel via cross-correlation with the template. This ensemble of ToAs was used to generate timing residuals and to fit a stable timing solution. \textsc{tempo2} is utilized to fit for various timing parameters, including position, proper motion, spin frequency and its derivatives, DM and its derivatives, and binary parameters for binary pulsars. This fitting process is typically repeated, until a stable timing solution is obtained. The two PTA sources in our sample, PSRs~J1022+1001 and J2145$-$0750, are usually provided with an already, relatively stable timing solution thanks to the European PTA Data Release 2~\citep{eptadr2}. 
Once, a reliable timing solution was obtained, the new ephemeris file was installed in the original observations, that are then time-averaged again. These archives were again used to make the final template repeating the procedure mentioned in \S~\ref{cretempl}.

\subsection{Obtaining DM time-series}
Using the noise-free template generated by the methods in \S~\ref{cretempl}, the \texttt{pat} module was employed to produce topocentric ToAs for each frequency channel. Following \cite{tiburzi2019} and \cite{donner2020}, a DM time-series was generated using an approach known as the ``Epoch-wise'' method~\citep{fra2024}. This technique fits the ToAs belonging to the same observing epoch to the functional form shown in Eq.~\ref{timedel} using \texttt{tempo2}~\citep{Hobbsetal2006a}, to obtain the residual DM induced by the added time delay. The result is a DM measurement per epoch. All data products—including the DM time series, timing solutions, and ToAs—were obtained concurrently using the methods described in this section. The resulting spin and orbital parameters for all pulsars are summarised in Table \ref{tab:postfit_table}.

\section{Results and Discussions}
\label{results}
This section outlines the properties of each pulsar. As already mentioned, PSRs~J1022+1011 and J2145$-$0750 are included in the samples of the European and International PTAs; they are old, recycled pulsars with periods of less than $20$~ms. Their stability and predictability make them ideal candidates for precision pulsar timing and for gravitational wave studies~\citep{verbiest2016,Perera2019}.

\begin{table}[!ht] 
    \centering
    \begin{tabular}{cccl} 
        \toprule
        Pulsar Name 
        & \multicolumn{2}{c}{DM Uncertainty ($\times 10^{-5}$ pc\ cm$^{-3}$)} \\
        (J2000) & Median & Minimum \\ 
        \midrule 
        J0034$-$0534 & 6 & 1 \\
        J0826+2637 & 19 & 5 \\
        J1022+1001 & 16 & 2 \\ 
        J1136+1551 & 13 & 3 \\ 
        J1300+1240 & 13 & 4 \\
        J1645$-$0317 & 9 & 3 \\
        J2145$-$0750 & 8 & 0.9 \\ 
        \bottomrule
    \end{tabular}
    \centering
    \caption{DM uncertainties for selected pulsars observed with I-LOFAR, showing the median and minimum uncertainties.}
    \label{tab:dm_uncert_table}
\end{table}

\subsection{DM timeseries}
\label{dmts_sec}
\begin{figure}[!h]
    \centering
    \includegraphics[width=1\linewidth,height=1.2\linewidth]{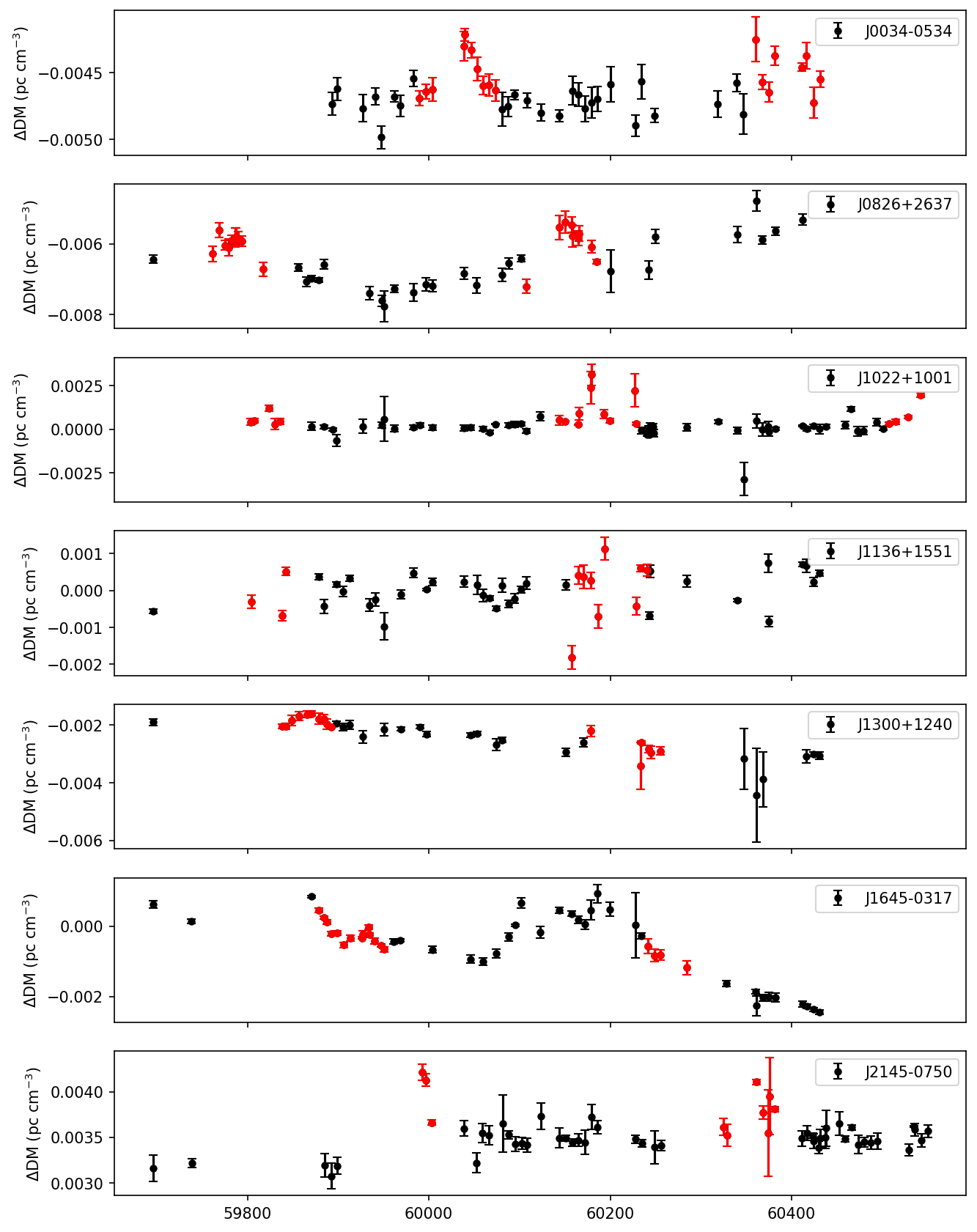}
    \caption{DM time series of each of the pulsars. The red points in each panel are the epochs that are less than 45~degrees away from the Sun.}
    \label{fig:dmts_all}
\end{figure}

The DM time series are displayed in Fig.~\ref{fig:dmts_all}, with red points marking epochs when the pulsar’s LoS comes within 45$^{\circ}$ of the Sun. Significant fluctuations in the DM time series are observed especially for PSRs~J0034$-$0534, J0826+2637, J1022+1001, and J2145$-$0750 as their LoS near the Sun. These emphasize their potential for studying SW effects, as their ecliptic latitudes make them ideal candidates for such investigations using pulsars.

Other pulsars, such as PSRs~J1300+1240 and J1645$-$0317, show gradual trends in their DM time-series, indicating potential variations in the IISM electron content along their line of sight. These gradual changes may reflect evolving electron densities within the IISM, highlighting the importance of studying long-term dispersion effects with low-frequency observations. Tab.~\ref{tab:dm_uncert_table} provides details of each pulsar along with their DM uncertainty. Some pulsars demonstrate exceptionally high DM precision, suggesting that our observations are approaching the sensitivity needed to detect ionospheric variations within the DM time-series. If we consider a total electron content of the ionosphere to be 10-60~TECU which is the normal value over a full day at Birr (see \cite{impc}\footnote{\texttt{https://impc.dlr.de}} for TECU maps), using Eq.~22 in \citet{lunaska}, we get a DM contribution of $3 \times 10^{-6} - 1.9\times 10^{-5}$~$\textrm{pc cm}^{-3}$. The high fractional bandwidth and enhanced sensitivity of future instruments like the Square Kilometer Array (SKA) \citep{braun_ska} will enable significantly improved precision, providing a promising avenue for detailed investigations of the ionosphere using pulsars.

\begin{figure*}[!htp]
    \centering
    \includegraphics[width=1\linewidth,height=0.95\linewidth]{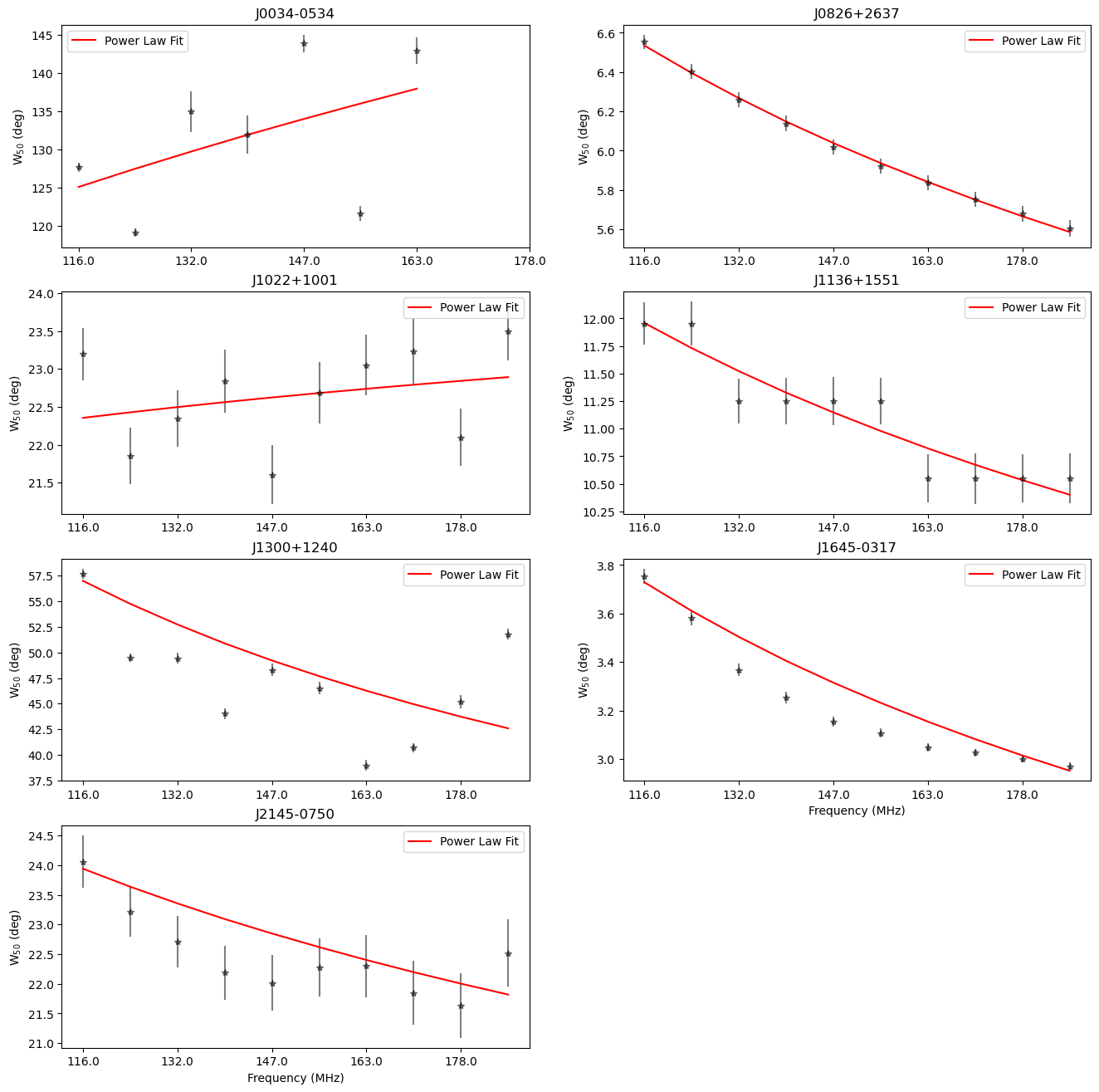}
    \caption{Full-width-half-maximum (FWHM) pulse widths for pulsars in our dataset. The black stars represent the pulse width at each frequency with the corresponding error in measurement. The red lines in each panel are the power law fits on the FWHM values. The FWHM values that lie below the power law fits correspond to the absorption features. It can be seem most prominently in PSRs~J1300+1240 \& J1645$-$0317.}
    \label{fig:fwhm_plot}
\end{figure*}

\subsection{Frequency Evolution of Pulsar Pulse Profiles}

Multi-frequency pulse profiles offer crucial information on pulsar emission geometries, including emission region heights, beam shapes, and radius-to-frequency mapping \citep{lyne1988,Rankin1983a,Rankin1983b,Mitra1999,Hankins2010,hassall2013}. A comprehensive study of 74 pulsars by \citet{Xu2021} provided significant information on the frequency evolution of pulse profiles. For 71 pulsars, the pulse width $W_{50}$ (full width of the pulse profile at 50\% of the peak profile amplitude, FWHM), follows a power-law trend at both high and low frequencies. However, deviations from this power-law behaviour were observed in the mid-frequency range (100 to 800~MHz), consistent with absorption characteristics proposed for the first time by \citet{Rankin1983b}. 

Following \citet{Xu2021} we use the FWHM as a metric to study the beam widths as a function of frequency of the pulsars in our sample at the exceptionally long wavelengths offered by I-LOFAR. We obtain $W_{50}$ by fitting multiple Gaussians, i.e.
\begin{equation}
    y=\sum_{i=1}^n A_i\exp\left(\frac{-(x-\mu_i)^2}{2\sigma_i^2}\right)\,,
    \label{gaussian}
\end{equation}
where $A_i,\mu_i\ \textrm{and}\ \sigma_i$ are amplitude, mean and standard deviation of the $i^{\rm th}$ Gaussian component. The $W_{50}$ and its uncertainty $W_{50,i}^{err}$ are then given by:
\begin{equation}
\begin{split}
    W_{50,i} = 2\sqrt{2\ln 2}\ \sigma_i\,, \\
    W_{50,i}^{err} = 2\sqrt{2\ln 2}\ \sigma_i^{\rm err}\,
\end{split}   
\end{equation}
where $\sigma_i^{err}$ is the uncertainty on the standard deviation of the Gaussian component. For pulsars with multiple peaks exceeding 50\% of the peak amplitude, where the intervening trough falls below this threshold, such as certain frequency profiles of PSR~J0034$-$0534~and all profiles in PSR~J1136+1551, we manually measured $W_{50}$ as the individual $W_{50,i}$ values or their sum did not accurately represent the beam width. This is the reason why the $W_{50}$ values for PSR~J1136+1551 appear to be quantised. To get the error on the FWHM, we used the quadrature sum of the $W_{50,i}^{\rm err}$ of the individual Gaussian components. The frequency evolution of these pulsars is fit with a power law function of the form $W_{50}\propto f^{\alpha}$.



\begin{figure*}[hb]
    \centering
    \includegraphics[width=1\linewidth,height=0.95\linewidth]{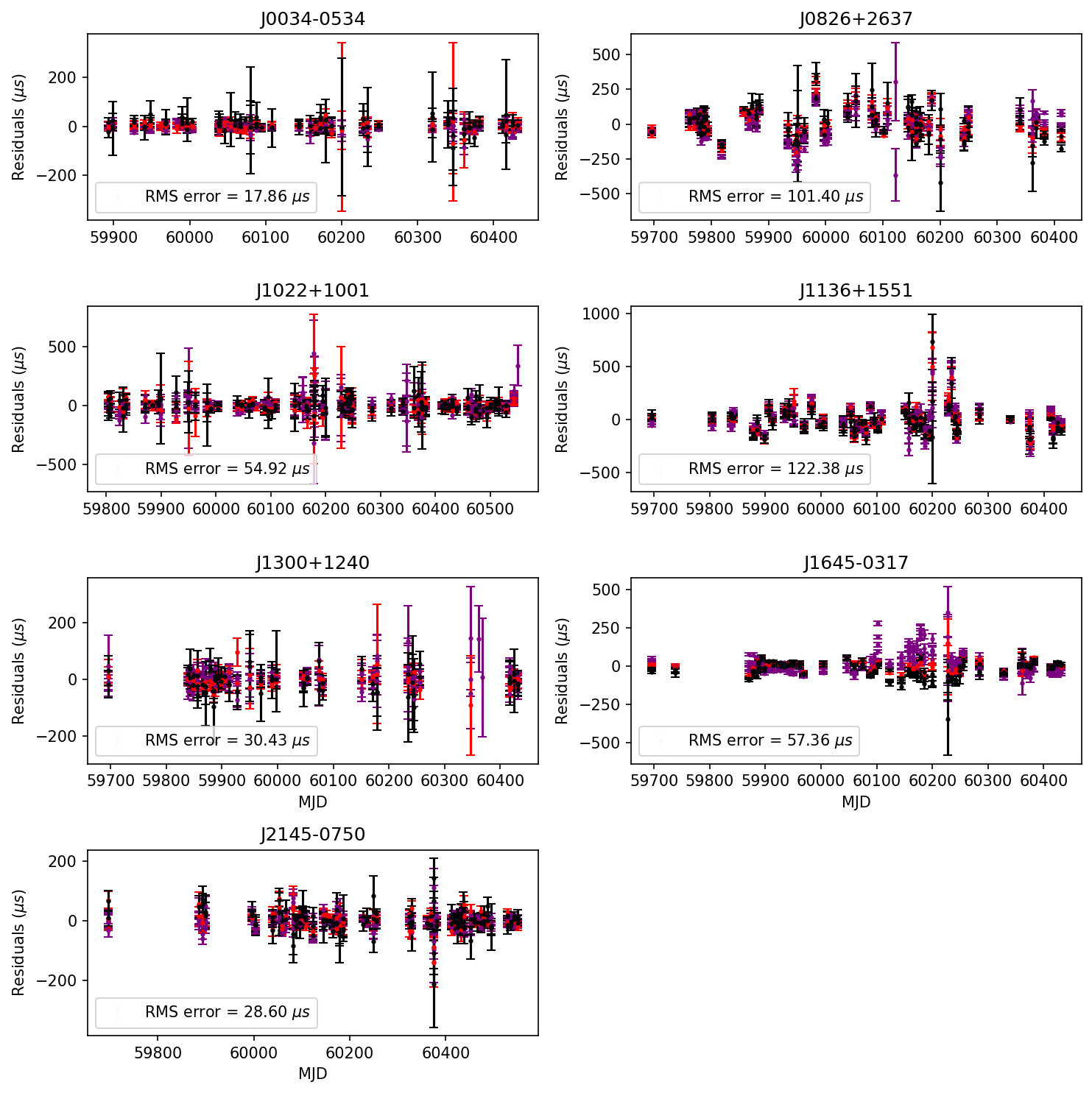}
    \caption{Residuals after timing using \texttt{tempo2}. The different colours are representative of different frequencies.}
    \label{fig:resids_all}
\end{figure*}

\begin{figure*}[hb]
    \centering
    \includegraphics[width=1\linewidth,height=0.5\linewidth]{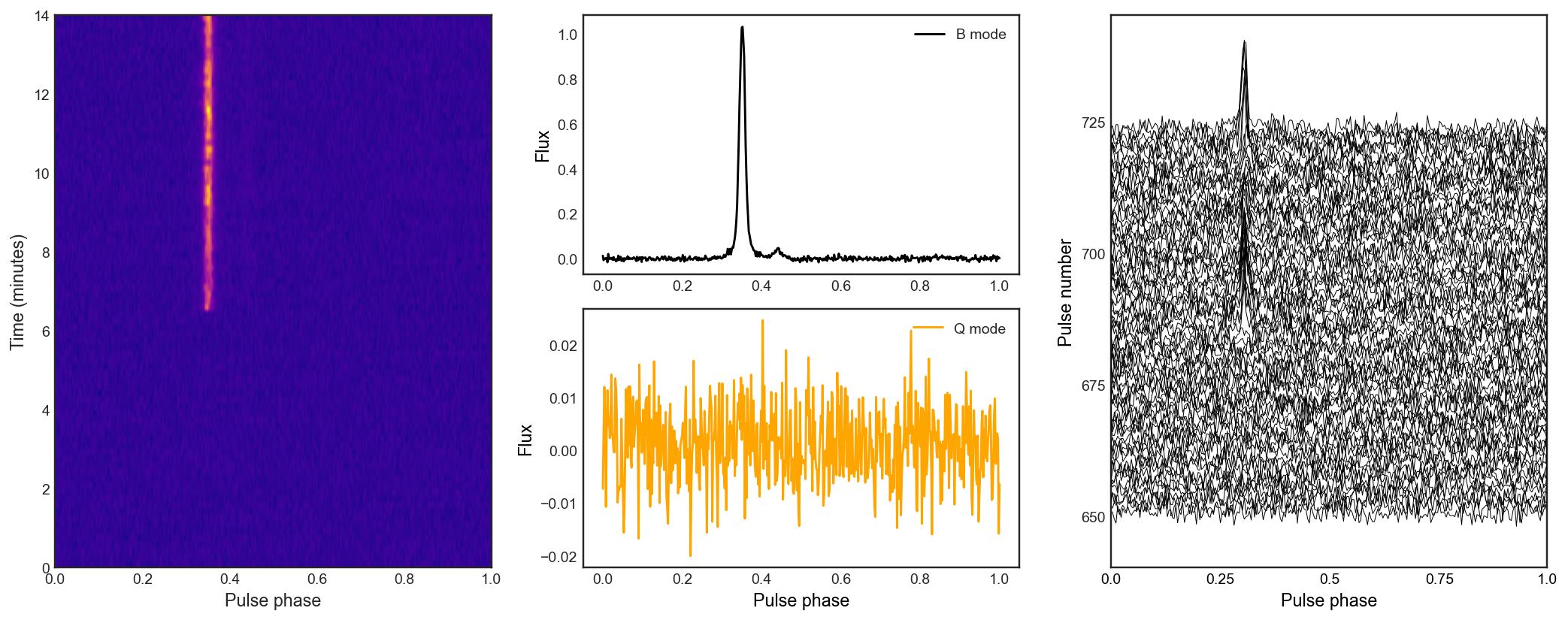}
    \caption{The observation of PSR~J0826+2637 from the observation on April 18th, 2023. The left column shows the emission as a function of time in 10-s sub-integrations;  the transition in emission modes is evident. The middle column shows the pulse profile during the B (Q) mode in the top (bottom) panel. The right column shows the `Joy-Division' plot, displaying rotation period numbers 650 to 800. The transition from the Q-mode to the B-mode is marked by a sudden increase in pulse brightness. }
    \label{fig:0826_nulling}
\end{figure*}

\begin{figure}[!h]
    \centering
    \includegraphics[width=1\linewidth,height=0.85\linewidth]{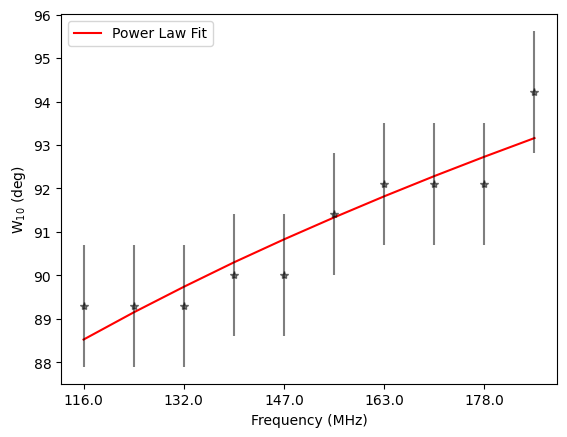}
    \caption{Full width at 10\% of the peak flux density ($W_{10}$) for PSR~J2145$-$0750 as a function of frequency. Unlike the $W_{50}$ values shown in Fig.~\ref{fig:fwhm_plot}, $W_{10}$ exhibits a clear increasing trend. The red line indicates a power-law fit to the data.}
    \label{fig: fwtm_2145}
\end{figure}

Fig.~\ref{fig:fwhm_plot} shows $W_{50}$ as a function of observing frequency. The red lines represent the power law fits on the $W_{50}$ values. We see that for some pulsars the $W_{50}$ values lie below the power law fit. These so-called ``absorption'' features, could pose challenges to the conventional pulsar radiation models \citep{Rankin1983b,Xu2021}. They are attributed to cyclotron absorption and it is likely magnetospheric in origin \citep{bartel1981}. Most pulsars exhibit a general trend of decreasing pulse width with increasing frequency. This is the trend expected from a radius-to-frequency mapping along primarily dipolar field lines where higher-frequency emission originates closer to the stellar surface \citep{ruderman1975}. However exceptions exist, notably, PSR~J0034-0534 where the $W_{50}$ values appear to increase as a function of frequency. Conversely, PSR~J1022+1001 shows minimal changes in $W_{50}$ across a wide frequency range, indicating stable profile morphology. PSR~J0826+2637 agrees with the findings of \citet{Rankin1983b} and \citet{Xu2021} , with a steady decrease in their pulse widths over the observed frequency range showing no noticeable absorption features. Whereas, PSRs~J1300+1240, J1645$-$0317 and J2145$-$0750 show absorption features in the LOFAR band itself. To address these exceptions, we also analysed the values of full width at 10\% of maximum ($W_{10}$) as, depending on the particle profile shape (i.e. on the viewing geometry), different width metrics might be more representative of the emission beam widths~\citep{lyne2010}. We saw similar trends across the entire frequency range for all sources, with the exception of PSR~J2145$-$0750 where its $W_{10}$ values seemed to be \textit{increasing} with frequency, contrary to its $W_{50}$ trend as shown in Fig. \ref{fig: fwtm_2145}. This might imply that emission from the centre and edges of the beam are not coming from the same height above the surface for the same frequency, i.e. radius to frequency mapping does not hold. 

These findings underscore the importance of having a high fractional bandwidth. SKA1-Low, which is currently under construction in Western Australia, has a 7:1 instantaneous fractional bandwidth~\citep{braun_ska} providing the coverage that should enable comprehensive measurements of these trends in the Southern pulsar population and provide unique constraints on pulsar radiation models. With the aforementioned observed behaviour, and as a very bright source in this band, PSR~J2145$-$0750 is an obvious target for SKA1-Low even in the very early array assemblies.

\subsection{PSR~J0034$-$0534}
PSR~J0034$-$0534, classified as a binary millisecond pulsar \citep{bailes1994}, is one of the most precisely timed pulsars at low frequencies. We report a DM of 13.764996~$\pm$~0.000015~$\textrm{pc cm}^{-3}$ with a remarkable median uncertainty of $7.9 \times 10^{-5}$~$\textrm{pc cm}^{-3}$ for all observations. This high precision in DM measurements allows it to serve as an excellent probe for detecting subtle DM variations due to effects such as the SW and the ionosphere. With a rotational period of 1.8~ms, J0034$-$0534 has the third shortest period known among pulsars. It orbits a white dwarf companion in a 1.6~day binary system.

It has a broad integrated pulse profile with a minimal off-pulse region at 150 MHz. This can be observed in the template profile shown in Fig.~\ref{fig:freq_evol}.
The timing residuals have an rms uncertainty of 18.54~$\upmu$s as shown in Fig.~\ref{fig:resids_all}. Notably, for this pulsar the upper three frequency channels were excluded from the analysis due to the complete absence of signal. This manual procedure was required, likely as a result of the source’s very steep spectrum. Additionally, the exceptional DM precision of J0034$-$0534 provides evidence for potential asymmetry in SW-induced DM variations around solar conjunction, as noted in previous studies \citep{tiburzi2019, susarla2024}. Further investigation, ideally involving daily observations around solar conjunction, would be valuable to determine the sensitivity of this pulsar to such effects.


\subsection{PSR~J0826+2637 (B0826+23)}
One of the earliest discovered pulsars, PSR~J0826+2637, was first identified by \citet{craft1968} with a rotation period of 530~ms. Using the I-LOFAR dataset, we report a DM of 19.504089~$\pm$~0.00009~$\textrm{pc cm}^{-3}$ with an rms error in timing residuals of 101.40~$\upmu$s. Although it is among the brightest known pulsars, it exhibits complex emission characteristics that have been investigated extensively in studies such as \citet{sobey2015} and \citet{Rankin2020}. 

This pulsar displays two primary emission states: the Bright, or "B" mode, and the Quiet, or "Q" mode. The Q mode is characterized by weak and sporadic emission. Additionally, \citet{sobey2015} suggests the possible existence of a third, precursor mode with even lower flux that occasionally drops to zero, potentially reflecting magnetospheric processes; however, confirming this hypothesis requires an extended observation window.

Fig \ref{fig:0826_nulling} shows the two modes observed during the observation on 18$^{\textrm{th}}$ April 2023. The pulsar appears to be very weak in the Q mode, as evidenced in many observations where the S/N has significantly dropped, in this example by a factor of at least $50$. The left column in Fig.~\ref{fig:0826_nulling} shows a waterfall-plot for individual pulses on the same date. This precisely shows the progression from the Q-mode to the B-mode. This poses a challenge, albeit tractable, for timing as one must be careful to account for this moding, i.e. only consider times when the pulsar is in the B mode. This switching behaviour is seen in the wider pulsar sample when observing sensitivity and cadence allow~\citep{lyne2010}. It appears to be due to magnetospheric switching, i.e. the entire pulsar magnetosphere switching from one stable configuration to another~\citep{timokhin2010}. While this is now widely observed~\citep{keane2013} it remains an open question as to why pulsars switch at all, and why they switch back and forth between stable, possibly quantised, states~\citep{cordes2013}. This source presents an excellent opportunity to study this phenomenon as it reliably switches in a tractable timescale. It is to be expected that the spin-down rate in each mode is different, as the volume of the magnetosphere and pulsar wind have changed but the relative time spent in each mode may not permit this to be measured for this source. 

The switch itself is seen here to take less than 1 rotation period. Its light cylinder distance is $\sim 25300$~km, $\sim 84$~ms light travel time. The dynamical timescale $1/\sqrt{G\rho}\sim 0.5$~ms meaning that such large-scale reconfigurations could conceivably occur this rapidly. Detailed full-polarisation single pulse studies at the transitions of this source may be able to pin down the location and scale sizes of the switches.

\subsection{PSR~J1022+1001}
PSR~J1022+1001 is another precisely-timed pulsar. Due to its exceptional stability and relatively low declination, all PTAs worldwide include this pulsar in their observations, making it one of the most thoroughly studied millisecond pulsars of all times. In our analysis, we report a DM of 10.25321~$\pm$~0.00002~$\textrm{pc cm}^{-3}$ with a median DM uncertainty of $1.6 \times 10^{-4}$~$\textrm{pc cm}^{-3}$. Additionally, we report an rms error on the timing residuals of 54.92~$\upmu$s. With an ecliptic latitude of $-0.06^{\circ}$, the LoS of this pulsar passes through the Sun. This proximity allows for the examination of SW effects on the pulsar’s signal, as previously investigated by studies such as \citet{Tiburzi2021, susarla2024}. Particularly, \citet{susarla2024} showed that this pulsar has a consistently higher values of electron density at 1 AU after studying a decade's LOFAR data with a reasonably high cadence. This phenomenon is attributed to the fact that the LoS of this pulsar crosses through the slow wind region of the SW, which is of higher electron density. They also showed the need for a variable model of solar wind for PTA sources like these to mitigate this noise effectively. 

\subsection{PSR~J1136+1551}
PSR~J1136+1551 (or B1133+16) is an isolated pulsar with a period of 1.18~s, characterized by a double-peaked profile and a high S/N. In our analysis, we report a DM of 4.84821~$\pm$~0.00007~$\textrm{pc cm}^{-3}$, with an rms timing residual error of 122.38~$\upmu$s. Previous studies have documented the occurrence of unusually intense single pulses, known as giant pulses \citep{GP1133}, which provide an opportunity to explore the pulsar's emission mechanisms in greater depth \citep{oswald1133}.

\subsection{PSR~J1300+1240 (B1257+12)}
PSR~J1300+1240 was the first pulsar found to host a planetary system, with three planets detected in orbit around it \citep{B1257_planetary}. This discovery was also the earliest evidence of extrasolar planets. Located approximately $710$~pc from the Sun, the J1300+1240 system is thought to have originated from a white dwarf merger. The presence of orbiting planets necessitates a complex timing model to account for the influences of all planetary bodies. In our analysis, we report a DM of 10.1588~$\pm$~0.0028~$\textrm{pc\ cm}^{-3}$, with an rms timing residual error of 30.43~$\upmu$s. Notably, the DM of this pulsar varies over extended timescales, indicating turbulence in the local ISM. Due to its low ecliptic latitude, effects of the SW are seen in I-LOFAR data, particularly during the first solar conjunction near MJD 59900, as shown in Fig.~\ref{fig:dmts_all}. 

\subsection{PSR~J1645$-$0317}
PSR~J1645$-$0317 is a bright pulsar in the LOFAR band with a period of $\sim$387~ms. It has large DM variations with a considerable change of 0.002~$\textrm{pc cm}^{-3}$ over a period of two years (see Fig.~\ref{fig:dmts_all}). 
In our analysis, we report a DM of 35.7416~$\pm$~0.0022~$\textrm{pc cm}^{-3}$ and an rms timing residual error of 57.36~$\upmu$s.

\begin{figure}
    \centering
    \includegraphics[width=1\linewidth,height=0.85\linewidth]{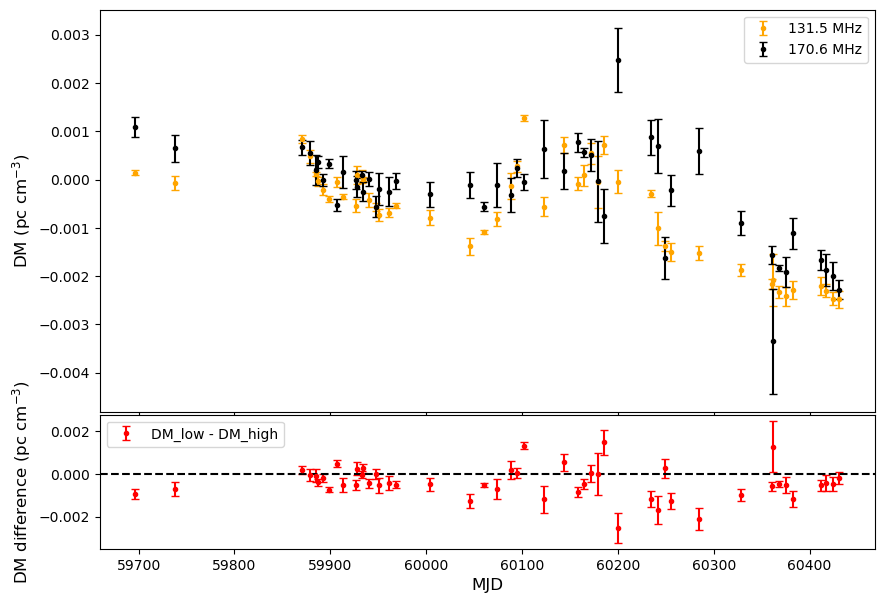}
    \caption{DM chromaticity in PSR~J1645$-$0317. In the top panel, the black points with errorbars show the DM timeseries at 170.6~MHz whereas the orange points show the DM timeseries at 131.5~MHz. The bottom panel shows the difference in the absolute values of their DMs at the same epoch. It is worth noting that the DM trends in both the bands are different highlighting the frequency-dependent DMs.}
    \label{dmchrom}
\end{figure}

Given the strong DM variations observed in this pulsar, we investigated the potential frequency dependence of DM~\citep{Cordes_2016}. To this end, we constructed a standard template comprising of 20 frequency channels using the methodology described in Sect. \ref{cretempl} and divided that into two templates of 10 frequency channels each. The observations were frequency-averaged into 20 channels and subsequently divided into two sub-bands, each consisting of 10 channels, with central frequencies of approximately 130~MHz for the lower sub-band and 170~MHz for the upper sub-band. Following the approach outlined in Sect. \ref{dmts_sec}, we generated DM time series for both sub-bands using their respective templates. The resulting DM time series, shown in Fig.~\ref{dmchrom}, exhibit systematic differences between DMs measured at the same epoch but in different frequency bands, providing tentative evidence for DM chromaticity. This effect has been previously reported for PSR~J2219+4754 \citep{donnerfddm2019} and is primarily attributed to small-scale, high-amplitude variations in the interstellar electron density. As discussed by \citet{Cordes_2016}, such variations arise from turbulence in the ionised interstellar medium, which induces frequency-dependent multipath propagation due to the resulting microstructure. While such chromatic DM variations could pose challenges for high-precision pulsar timing experiments, the long-term trends in both sub-bands appear to be consistent. Consequently, the impact on timing precision may be negligible.

\begin{table*}[!hb]
\centering
\renewcommand{\arraystretch}{1.2}
\caption{Spin and orbital parameters of the seven pulsars in our dataset.}
\begin{tabular}{lccccccccc}
\hline
Pulsar & RAJ & DECJ & F0 & F1 & F2 & DM & $P_b$ & $\dot{P_b}$ & A1 \\
(J2000) & (hh:mm:ss) & (dd:mm:ss) & (Hz) & ($10^{-15}$ s$^{-2}$) & ($10^{-26}$ s$^{-3}$) & (pc cm$^{-3}$) & (days) & ($10^{-11}$) & (lt-s) \\
\hline
J0034$-$0534 & 00:34:21.84 & $-$05:34:36.95 & 532.7134 & $-1.3484$ & -- & 13.765 & 1.5893 & -- & 1.4378 \\
J0826+2637 & 08:26:51.16 & +26:37:11.37 & 1.8844 & $-5.8973$ & $-7.2884$ & 19.504 & -- & -- & -- \\
J1022+1001 & 10:22:57.98 & +10:01:52.92 & 60.7794 & $-0.1608$ & -- & 10.253 & 7.8051 & 7.1432 & 16.7648 \\
J1136+1551 & 11:36:03.15 & +15:51:18.26 & 0.8418 & $-2.6224$ & $-10.472$ & 4.848 & -- & -- & -- \\
J1300+1240 & 13:00:03.12 & +12:40:54.41 & 160.8097 & $-2.9448$ & -- & 10.152 & 25.3168 & -- & 8.98$\times 10^{-6}$ \\
J1645$-$0317 & 16:45:01.93 & $-$03:18:09.26 & 2.5793 & 738.34 & -- & 35.741 & -- & -- & -- \\
J2145$-$0750 & 21:45:50.46 & $-$07:50:18.53 & 62.2959 & $-0.1184$ & -- & 9.003 & 6.8389 & 4.2981 & 10.1647 \\
\hline
\end{tabular}
\vspace{0.2cm}
\tablefoot{Pulsar names are in J2000 format. RAJ and DECJ are Right Ascension and Declination. F0 is the spin frequency; F1 and F2 are its first and second derivatives. DM is the dispersion measure. $P_b$ and $\dot{P_b}$ are the orbital period and its derivative, and A1 is the projected semi-major axis.}
\label{tab:postfit_table}
\end{table*}

\subsection{PSR~J2145$-$0750}
PSR~J2145$-$0750 is a bright pulsar with a period of 16 ms. In our analysis, we report a DM of 9.00272~$\pm$~0.00001~$\textrm{pc cm}^{-3}$ and an rms timing residual error of 28.60~$\upmu$s. With an ecliptic latitude of $\sim5^{\circ}$, it is an ideal probe to study the effects of SW. Due to its relatively stable timing solution \citep{verbiest2009}, PSR~J2145$-$0750 is also monitored as a PTA pulsar. Its low declination allows observation by every major PTA, providing a wealth of data for precision timing projects.

\section{Conclusions}
\label{conclusions}
This study presents long-term timing results of seven ecliptic pulsars observed with I-LOFAR over a two-to-three year period, focusing on the effects of dispersion due to the IISM and the solar wind. A significant outcome of this research are the precise measurements of DM using pulsar timing. The DM time series of four pulsars exhibit clear variations correlated with their ecliptic latitudes, highlighting the impact of the SW. We observe that a spherically symmetric electron density model of the SW is not able to completely model the variability that can be seen in Fig \ref{fig:resids_all} for pulsars like PSR~J1022+1001. The precise DMs obtained in this study emphasize the role of low-frequency observations in complementing higher-frequency PTA datasets (with much poorer DM precision), particularly in mitigating DM noise and SW noise. In gravitational wave studies these noise sources act as foregrounds \citep{verbiest2018}. The precision is also at the point that it may be possible to detect the variation in the ionospheric DM using only pulsar timing. The LOFAR DM precision is $\sim 30$~TECU and other instruments coming online under more active regions of the ionosphere should have better resolution still.

This study also examined the frequency evolution of pulse profiles, shedding light on pulsar emission geometries. While most pulsars seem to follow a consistent decrease in $W_{50}$ with increasing frequency, deviations from this, the expected behaviour, were observed. Separately, we examined the pulse-nulling phenomenon in PSR~J0826+2637. This pulsar shows extreme weakness in its Q-mode, with a sudden switch to the B-mode occurring within a single rotation. Such rapid mode transitions are likely magnetospheric in origin, suggesting quantized stable magnetospheric states and we see that the switching process occurs in less than one rotation period.

Furthermore, we investigated the DM chromaticity of PSR~J1645$-$0317. Our findings present tentative evidence for frequency-dependent dispersion effects in this source. To confirm this phenomenon robustly, dedicated observations with exceptionally high fractional bandwidth and sensitivity at low frequencies are essential. This could come from a combined campaign using, for example, LOFAR and the Giant Metrewave Radio Telescope, but SKA1-Low’s advanced capabilities are probably best suited for such studies, promising significantly improved DM precision. By utilizing bright pulsars like PSR~J1645$-$0317, even during commissioning, SKA1-Low~\citep{braun_ska} can serve as a critical tool for testing these effects, ultimately enabling more accurate pulsar timing and deepening our understanding of interstellar propagation.



\begin{acknowledgements}
SCS acknowledges the support of a University of Galway, College of Science and Engineering fellowship in supporting this work. OAJ acknowledges the support of Breakthrough Listen which is managed by the Breakthrough Prize Foundation. JPWV acknowledges support from the National Science Foundation (NSF) AccelNet program, award No. 2114721. The authors thank Letizia Vincetti for her valuable comments on the paper. The Rosse Observatory is operated by Trinity College Dublin. I-LOFAR infrastructure has benefited from funding from Science Foundation Ireland, a predecessor of Taighde \'{E}ireann --- Research Ireland.  
\end{acknowledgements}

\section*{Data Availability}
All the data presented are publicly available on Zenodo \citep{susarla_2025_dr}. This repository includes down-sampled archives, timing solutions, ToAs, templates and DM time series. Additional data can be provided upon request.



\bibliographystyle{aa.bst}
\bibliography{ie613} 


\label{lastpage}
\end{document}